# Solar Models with Helium and Heavy Element Diffusion

John N. Bahcall

School of Natural Sciences, Institute for Advanced Study, Princeton, NJ 08540

M. H. Pinsonneault

Department of Astronomy, Ohio State University, Columbus, OH 43210

With an Appendix on the Age of the Sun by G. J. Wasserburg

Division of Geological and Planetary Sciences, MS170-25

California Institute of Technology, Pasadena, CA 91125

## Abstract

Heavy element and helium diffusion are both included for the first time in this series of papers on precise solar models. In addition, improvements in the input data for solar interior models are described for nuclear reaction rates, the solar luminosity, the solar age, heavy element abundances, radiative opacities, helium and metal diffusion rates, and neutrino interaction cross sections. The effects on the neutrino fluxes of each change in the input physics are evaluated separately by constructing a series of solar models with one additional improvement added at each stage. The effective $1\sigma$ uncertainties in the individual input quantities are estimated and used to evaluate the uncertainties in the calculated neutrino fluxes and the calculated event rates for solar neutrino experiments.

The calculated neutrino event rates, including all of the improvements, are $9.3^{+1.2}_{-1.4}$ SNU for the $^{37}$Cl experiment and $137^{+8}_{-7}$ SNU for the $^{71}$Ga experiments. The calculated flux of $^7$Be neutrinos is $5.1(1.00^{+0.06}_{-0.07}) \times 10^9$ cm$^{-2}$s$^{-1}$ and the flux of $^8$B neutrinos is $6.5(1.00^{+0.14}_{-0.17}) \times 10^6$ cm$^{-2}$s$^{-1}$. The primordial helium abundance found for this model is $Y = 0.278$. The present-day surface abundance of the model is $Y_s = 0.247$, in agreement with the helioseismological measurement of $Y_s = 0.242 \pm 0.003$ determined by Hernandez and Christensen-Dalsgaard (1994). The computed depth of the convective zone is $R = 0.712\ R_\odot$ in agreement with the observed value determined from $p$-mode oscillation data of $R = 0.713 \pm 0.003\ R_\odot$ found by Christensen-Dalsgaard *et al.* (1991). Although the present results increase the predicted event rate in the four operating solar neutrino experiments by almost $1\sigma$ (theoretical uncertainty), they only slightly increase the difficulty of explaining the existing experiments with standard physics (i.e., by assuming that nothing happens to the neutrinos after they are created in the center of the sun).

For an extreme model in which all diffusion (helium and heavy element diffusion) is neglected, the event rates are $7.0^{+0.9}_{-1.0}$ SNU for the $^{37}$Cl experiment and $126^{+6}_{-6}$ SNU for the $^{71}$Ga experiments, while the $^7$Be and $^8$B neutrino fluxes are, respectively, $4.5(1.00^{+0.06}_{-0.07}) \times 10^9$ cm$^{-2}$s$^{-1}$ and $4.9(1.00^{+0.14}_{-0.17}) \times 10^6$ cm$^{-2}$s$^{-1}$. For the no-diffusion model, the computed value of the depth of the convective zone is $R = 0.726\ R_\odot$, which disagrees with the observed helioseismological value. The calculated surface abundance of helium, $Y_s = 0.268$, is also in disagreement with the $p-$mode measurement. We conclude that helioseismology provides strong evidence for element diffusion and therefore for the somewhat larger solar neutrino event rates calculated in this paper.

## CONTENTS

I. INTRODUCTION

Solar interior models have steadily improved in precision over the past three decades, in part due to the stimulus of solar neutrino experiments that are in apparent conflict with the model calculations. The interest in the details of these models has increased as many authors have suggested that the explanation between the apparent conflict between solar model calculations and solar neutrino observations may be due to new neutrino physics. The probability of transitions from the more easily detectable electron-type neutrinos to the more difficult to observe other types of neutrinos (muon and tau neutrinos) depends, in some models of particle physics, on characteristics of the solar interior (see, e.g., Mikheyev and Smirnov, 1986; Wolfenstein, 1978; Gribov and Pontecorvo, 1969; Roulet, 1991; Guzzo, Masiero, and Petcov, 1991; Lim and Marciano, 1988; Akhmedov, 1988).

Helioseismological analysis of measured low-$l$ $p$-mode frequencies provide increasingly accurate tests of the sound velocity in the solar interior and of the depth of the convective zone. Unlike the neutrino observations (Hirata *et al.*, 1991; Davis 1993, Anselmann *et al.*, 1994; Abdurashitov *et al.*, 1994), the measured $p$-mode frequencies are in good quantitative agreement (typically to much better than one part in a thousand) with the most accurate solar model calculations, especially when the theoretical uncertainties are taken into account. The agreement between $p-$mode frequencies and solar model calculations has been achieved as a result of iterations involving successive improvements in both the theory and the observations. One of the earliest achievements of helioseismology led to more precise calculations of the depth of the solar convection zone, which now agrees well with the helioseismological determination (for discussions of the observed and calculated solar $p$-mode oscillations see, e.g., Libbrecht, 1988; Bahcall and Ulrich, 1988; Elsworth *et al.*, 1990; Gough and Toomre, 1991; Christensen-Dalsgaard, Gough, and Thompson, 1991; Guenther, Pinsonneault, and Bahcall, 1993; Dziembowski, Goode, Pamyatanykh, and Sienkiewicz, 1994). In the present paper, we show that a recent helioseismological determination of the present-day surface abundance of helium by Hernandez and Christensen-Dalsgaard (1994) agrees with the results of solar model calculations that include diffusion but disagrees with solar model calculations that do not include diffusion.

The neutrino fluxes calculated using different solar model codes are in good agreement with each other when the same input parameters are adopted. Typically, the agreement is better than or of the order 2% (Bahcall and Pinsonneault, 1992; Bahcall and Glasner, 1994; Bahcall, 1994). Recently published calculations of standard solar models and solar neutrino fluxes include work by Bahcall and Ulrich (1988), Turck Chièze et al. (1988), Sackman et al. (1990), Bahcall and Pinsonneault (1992), Turck Chièze and Lopes (1993), Castellani et al. (1994), Kovetz and Shaviv (1994), Christensen-Dalsgaard (1994), Shi et al. (1994), and Bahcall and Glasner (1994). Earlier models in the present series of papers, which began in 1963 (Bahcall, Fowler, Iben, and Sears, 1963), are described in Bahcall (1989).

One of the principal improvements that has been made in recent years is to include in the calculations the effects of element diffusion. In the absence of an external field, diffusion smooths out variations. However, in the case of the sun, the stronger pull of gravity on helium and the heavier elements causes them to diffuse slowly downward (toward the solar interior) relative to hydrogen. In a previous paper (Bahcall and Pinsonneault, 1992), we incorporated an approximate analytic description (Bahcall and Loeb, 1990) of the effects of hydrogen and helium diffusion into our stellar evolution code and evaluated the effects of this diffusion on the calculated neutrino fluxes, the depth of the solar convective zone, and the primordial helium abundance. In a related paper (Guenther, Pinsonneault, and Bahcall, 1993), the effects of helium and hydrogen diffusion on the calculated $p$-mode oscillation frequencies were evaluated using the same (Bahcall-Loeb) approximate analytic treatment of diffusion. The inclusion of helium and hydrogen diffusion somewhat exacerbated the differences between calculated and observed neutrino rates while improving the agreement with the $p$-mode oscillations (cf. Christensen-Dalsgaard, Proffitt, and Thompson, 1993).

A more accurate numerical solution of the fundamental equations of diffusion has subsequently been carried out by Thoul, Bahcall, and Loeb (1994) (hereafter TBL). The principal difference between the work of Thoul et al. and earlier studies (such as the well-known in-vestigation by Michaud and Proffitt 1992) is that Thoul et al. solves the Burgers equations exactly and then represents the numerical results by simple analytic functions, rather than trying to obtain analytic solutions by approximations. The results of TBL are available in a convenient exportable subroutine that can be included in solar model calculations. The TBL subroutine includes heavy element, helium, and hydrogen diffusion. The number of heavy elements that are diffused by this subroutine can be chosen by the user. In the present paper, we carry out the first evaluation of heavy-element as well as hydrogen and helium diffusion using the results of Thoul and her collaborators. We also take this occasion to update the input data for the nuclear reactions, the solar age, the element abundances, and the radiative opacities. Previous calculations including the effects of heavy element diffusion have been carried out by Proffitt (1994) and by Kovetz and Shaviv (1994). Although different formulations of the diffusion equations were used in the Proffitt (1994), the Kovetz and Shaviv (1994), and the present calculations, and somewhat different input data were adopted in each case, similar effects are found in all three calculations.

The primary goal of this paper is to provide accurate predictions, with well-defined uncertainties, of the solar neutrino fluxes in order to compare the expectations based upon standard physics (standard solar models and standard electroweak theory) with solar neutrino experiments. In order to orient the reader, we summarize below the results from the four operating solar neutrino experiments (where 1 SNU equals $10^{-36}$ interactions per target atom per sec) and give a preview of the theoretical results we obtain in this paper.

The measured rate for the chlorine experiment (which is primarily sensitive to $^8$B and $^7$Be neutrinos) is (Davis 1993, Cleveland et al. 1995):

$$\text{Rate(chlorine)} = 2.55 \pm 0.17(\text{stat}) \pm 0.18(\text{syst}). \tag{1}$$

We shall show in this paper that the standard model prediction is $9.3^{+1.3}_{-1.4}$ SNU including metal and helium diffusion and $7.0^{+0.9}_{-1.0}$ SNU without any diffusion. The measured rate for the GALLEX and the SAGE gallium solar neutrino experiments (which are primarily sensitive to $p-p$ and $^7$Be neutrinos) are, respectively (Anselmann et al. 1993,1994):

$$\text{Rate(gallium)} = 79 \pm 10(\text{stat}) \pm 6(\text{syst}), \tag{2}$$

and (Abdurashitov et al. 1994, Nico et al. 1995):

$$\text{Rate(gallium)} = 69 \pm 10(\text{stat}) \pm 6(\text{syst}). \tag{3}$$

For comparison, the standard model prediction for a gallium experiment is $137^{+6}_{-7}$ SNU including diffusion and $126^{+6}_{-6}$ SNU without diffusion. The measured flux (above 7.5 MeV) of $^8$B neutrinos found in the Kamiokande water-Cherenkov experiment is (Suzuki et al.1995, Hirata et al. 1991):

$$\text{Rate(water)} = [3.0 \pm 0.41(\text{stat}) \pm 0.35(\text{syst})] \times 10^6 \text{ cm}^{-2}\text{s}^{-1}. \tag{4}$$

The observed flux of $^8$B neutrinos is 0.45 of the standard model rate including diffusion and 0.61 of the standard model without diffusion.

The present paper is organized as follows. In § II, we describe improved input data and compare with the parameters that were used in earlier solar model calculations. In § III, we describe and compare the various prescriptions for element diffusion. We present our principal results on a series of solar models in § IV. We describe in § V how we calculate uncertainties in the predicted fluxes and event rates and present in this section our best estimates of the uncertainties caused by each of the most important parameters. We summarize and discuss our main conclusions in § VI.

II. INPUT DATA

We review in this section some of the important data that are used in constructing solar neutrino models, emphasizing the improvements that have been made since our last systematic investigation (Bahcall and Pinsonneault, 1992) and the error estimates for the different parameters. In § II.A, we review recent progress in determining nuclear reaction rates and present a summary table of the cross-section factors and uncertainties we adopt. We evaluate in § II.B the total solar luminosity by averaging the results of several different satellite measurements over the solar cycle. We analyze in § II.C and in Appendix A the meteoritic constraints on the solar age and determine a best-estimate, with uncertainties, for the age of the sun. We summarize in § II.D the current best estimates for the individual heavy element abundances and compare the present estimates with values determined over the past two decades. We describe in § II.E the OPAL radiative opacities used in our solar models. Finally, we present in § II.F the improved neutrino interaction cross sections used in the present paper.

A. Nuclear Reaction Rates

The principal progress on the nuclear reaction rates since 1992 has been theoretical, including a recalculation of the nuclear matrix element for the $p$-$p$ reaction (Kamionkowski and Bahcall, 1994a) and a self-consistent evaluation of the effects of vacuum polarization on the rates of the other important solar nuclear reactions (Kamionkowski and Bahcall, 1994b). Two recent reviews summarize the experimental situation (Parker, 1994) and the theoretical situation (Langanke, 1994) and discuss the validity of different ways of obtaining the extrapolated cross section factors. (In § VI, we give predicted event rates for two extreme assumptions regarding how the average experimental $^7\text{Be}(p,\gamma)^8\text{B}$ cross section is calculated. The two assumptions give predicted event rates within the quoted $1\sigma$ overall uncertainties.)

Table I gives in column five the nuclear reaction rates used here and compares those rates with the earlier values, listed in column four, used in Bahcall and Pinsonneault (1992). We also include in column three of Table I references to some of the recent papers on individual reactions; column six contains, where appropriate, explanatory comments on the reactions.

B. Solar Luminosity

In this subsection, we use the results from a series of recent satellite measurements to determine a best-estimate solar luminosity with an approximate uncertainty that spans all of the recent determinations of the luminosity.

The absolute luminosity of the sun is known with less accuracy than one might imagine; in fact, the recognized experimental uncertainties exceed the errors that are often quoted in the astrophysical literature. The principal experimental complications are: 1) the difficulties in making comparisons between the absolute sensitivities of different satellite radiometers; 2) the variability of the solar luminosity over the solar cycle; and 3) the systematic uncertainties associated with long-term solar variability (such as the Maunder minimum in the 16th and 17th century). In the absence of a theoretical understanding of the long-term variability, we do not make here an explicit calculation of its contribution to the overall uncertainty, although conventional wisdom suggests that long-term variability would not be as large an effect as the existing dispersion in absolute measurements between different satellite experiments.

In recent years, there have been a number of analyses of measurements of the total solar irradiance that were made with the aid of space-born electrically-self-calibrating cavities. Precise measurements have been carried out on a variety of satellites during the solar cycles 21 and 22, beginning in the year 1978 with the Earth Radiation Budget experiment launched on the Nimbus 7 spacecraft (see, e.g., Hickey *et al.*, 1980). Other important experiments include the ACRIM I detector on the Solar Maximum Mission (Willson *et al.*, 1981) and the Earth Radiation Budget Experiment instruments on the Earth Radiation Budget Satellite and the NOAA9 and NOAA10 Satellites (e.g., Hickey *et al.*, 1982). Measurements are also available from the second-generation ACRIM II experiment on the Upper Atmosphere Research Satellite, launched in September 1991 (see, e.g., Willson, 1993a). The precise relative measurements as a function of time from these experiments are in excellent agreement with each other and reveal a systematic, peak-to-peak variation with epoch in the solar cycle of order $\pm 0.1\%$ of the total irradiance (see, e.g., Lee *et al.*, 1991; Chapman *et al.*, 1992; Fröchlich, 1992; Hoyt *et al.*, 1992; Willson, 1993a). The average ratio of absolute irradiances measured by the different experiments on these satellites has a total dispersion of about $\pm 0.35\%$ (cf. Willson, 1993a, Table 1).

We have integrated the 81 day running means (Frochlich, 1992) of the solar irradiance from November 1978 to January 1991, adopting the absolute calibration given by the ACRIM experiment. The result for the absolute luminosity is

$$L_\odot = (1367 \pm 5) \text{ J m}^{-2} = 3.844(1 \pm 0.004) \times 10^{33} \text{erg s}^{-1}. \tag{5}$$

Willson (1993b) calculates a weighted mean average of $(1367.2 \pm 0.01)$ J m$^{-2}$ for all of the ACRIM I results, in good agreement with the above value.

The most important aspect of Equation (5) for our purposes is the systematic uncertainty in the absolute value, which dominates the error estimate. The uncertainty that we have adopted spans the range of measurements of the total irradiance obtained with different satellite radiometers during solar cycles 21 and 22. The previous value of the solar luminosity used in this series was (Bahcall *et al.*, 1982) determined from the early satellite measurements made in 1980 and 1981 and was $L_\odot = 3.86(1 \pm 0.005) \times 10^{33}$erg s$^{-1}$, which is 0.4% higher than the currently-recommended value.

Since the best-estimate value of $L_\odot$ has changed over the past decade by an amount equal to the estimated uncertainty shown in Equation (5), we choose to regard the quoted error, $\pm 0.4\%$, as a $1\sigma$ effective uncertainty. The estimated uncertainty in the luminosity of the sun corresponds (Bahcall, 1989) to less than a $\pm 3\%$ ($1\sigma$) uncertainty in all the solar neutrino fluxes.

C. Solar Age

The solar age is relatively well-determined from meteoritic measurements. A systematic analysis of the current state of our knowledge is given in Appendix A, which was written by G. J. Wasserburg. The best-estimate value is

$$t_{\odot \text{ age}} = (4.57 \pm 0.02) \times 10^9 \text{yr}, \tag{6}$$

where the quoted uncertainty includes errors of a systematic character. The previous value of the solar age used in this series was (Bahcall *et al.*, 1982) $t_{\odot \text{ age}} = (4.55 \pm 0.1) \times 10^9$ yr, which was based upon earlier studies of Wasserburg *et al.* (1977) and Wasserburg,

Papanastassiou, and Lee (1980). Tilton (1988) estimates the age of meteoritic condensation to be $4.56 \times 10^9$ yr. Guenther (1989), without access to all of the meteoritic information, recommended $4.49 \times 10^9$ yr.

Conservatively, we take the error estimate shown in Equation (6) to be an effective $2\sigma$ uncertainty (cf. Appendix A). The estimated uncertainty of about 0.5% in the age of the sun corresponds (Bahcall, 1989) to less than a 1% ($1\sigma$) uncertainty in all the solar neutrino fluxes.

A small fraction, less than 1%, of the estimated solar age is likely spent on pre-main sequence evolution (see, e.g., Iben and Talbot, 1966). Most modern solar models give a pre-main sequence lifetime before reaching the ZAMS of only about $3 \times 10^7$ yr.

All of the solar models constructed in this paper begin on the zero-age sequence. Different assumed pre-main sequence scenarios affect the predicted neutrino fluxes by less than 1% (see Bahcall and Glasner, 1994).

D. Element Abundances

The assumed relative abundances of the heavy elements in the primordial sun influence significantly the calculated neutrino fluxes (see Bahcall and Ulrich, 1971 for an early discussion). The heavy elements contribute importantly to the calculated radiative opacities, which in turn affect the temperature gradient in the solar interior model and therefore the neutrino fluxes. In addition, the assumed heavy element abundances affect the calculated mean molecular weight and therefore the stellar structure and neutrino fluxes.

Two different sources are predominantly used to determine primordial element abundances, the meteorites (carbonaceous CI chondrites) and the solar photosphere. Over the past several decades, there have been many initial disagreements between abundances determined from meteorites and abundances determined from the solar atmosphere. In nearly all cases, as the atomic data were steadily improved and the measured made more precise, the atmospheric values approached the meteoritic values. This trend has resulted in a consensus view that the meteoritic values correctly represent the primeval solar abundances (see, e.g., Grevesse and Noels, 1993a,b). Most importantly, the recent photospheric iron abundances are now in good agreement with the meteoritic values.

Table II shows the most recent revision by Grevesse and Noels (1993a) of the element abundances. We adopt their results for our best solar models. Also shown for comparison in Table II are the element abundances of Anders and Grevesse (1989), which were used in our previous best solar models (Bahcall and Pinsonneault, 1992), the element abundances determined by Grevesse (1984), which were used in the Monte Carlo simulations of Bahcall and Ulrich (1988), the element abundances of Ross and Aller (1976), which were used for many of the earlier papers in this series (beginning with Bahcall *et al.*, 1982), and the Lambert and Warner abundances (Lambert and Warner, 1968), which were used in the early discussion of Bahcall and Ulrich (1971).

It is instructive to compare in Table II the numerical values of the relative abundances that have been determined over the past quarter of a century. There is good agreement among the various determinations of nearly all of the heavy element abundances that are important for solar neutrino calculations. Judging by the trend indicated in Table II, the errors quoted by the various investigators appear to be reliable indicators of the actual uncertainties. Neon may be the most problematic of the heavy elements, since it is estimated to be rather abundant but it cannot be measured directly in the sun. The effect of the assumed neon abundance on the solar neutrino fluxes will be investigated separately by one of us (MHP).

How uncertain are the primordial abundances of the heavy elements? This is a difficult question to answer unambiguously since the most important uncertainties are probably systematic. If we adopt as the estimator of the likely uncertainty the variation of the measured value of $Z/X$ over the past decade, from Grevesse (1984) to Grevesse and Noels (1993a), we find a $1\sigma$ error of

$$\sigma(Z/X) = \pm 0.061 \times (Z/X), \tag{7}$$

which is within a few percent of the uncertainty estimated by Bahcall and Ulrich (1988). We note in passing that a given fractional change in the CNO abundances is much less important for calculating the neutrino fluxes than the same fractional change in the abundances of the heavier elements like Fe (see, e. g., Paper I).

E. Radiative Opacities

The improved radiative opacities (Iglesias and Rogers, 1991; Rogers and Iglesias, 1992, and references cited therein) computed by the OPAL group at the Lawrence Livermore National Laboratory have helped to resolve several long-standing discrepancies between observations and calculations in the field of stellar evolution and stellar structure (see Rogers and Iglesias, 1994 for an excellent review of the recent progress). For all of the solar models discussed in this paper, we have used OPAL opacities. Our best solar model was computed using opacities generously prepared by Iglesias and Rogers for the most recent determination of heavy element abundances by Grevesse and Noels (1993a) (see column two of Table II).

For uncertainties in solar opacities, we use the estimates that are given in Bahcall and Pinsonneault (1992). These estimates were derived by comparing the values obtained for each neutrino flux from solar models computed with the older Los Alamos opacities and with the more accurate OPAL opacities (see Equation 20 and the discussion in § VIII of Bahcall and Pinsonneault, 1992).

The ratio of mixing length to pressure scale height, $\alpha$, that reproduces the solar radius in a solar model calculation is strongly influenced by the low-temperature opacities and also by the choice of a model atmosphere; neither the low-temperature opacities nor the choice of a model atmosphere has a significant effect on the calculated neutrino fluxes. For completeness, we note that we used the Kurucz (1991) low-temperature opacities and the Krishna Swamy (1966) empirical solar $T - \tau$ relationship (in place of the previously used Eddington gray atmosphere). These choices for the input physics improve somewhat the agreement between the calculated and the observed $p-$mode oscillation frequencies (Guen-ther *et al.*, 1992) and lead to larger values of the mixing length. A more detailed discussion of the relevant outer-layer physics used can be found in Guenther *et al.* (1992).

F. Neutrino Interaction Cross Sections

For nearly all the neutrino interaction cross sections, we use the values calculated by Bahcall and Ulrich (1988) and tabulated in Bahcall (1989). The exceptions are: the cross sections for neutrino absorption by the high-energy $^8$B neutrinos incident on $^{37}$Cl, and the neutrino-electron scattering cross sections. In both cases, the changes are small.

For the $^8$B neutrino absorption cross section on $^{37}$Cl, we use the recent determination by Aufderheide *et al.* (1994) of $\sigma(^8\text{B}) = (1.11 \pm 0.08) \times 10^{-42}$ cm$^2$ ($3\sigma$ uncertainty), which is in good agreement with the Bahcall and Holstein (1986) value (adopted by Bahcall and Ulrich, 1988) of $\sigma(^8\text{B}) = (1.06 \pm 0.10) \times 10^{-42}$ cm$^2$. The slightly greater accuracy of the Aufderheide *et al.* value is the result of improved measurements of the $\beta$−decay strengths in the decay of $^{37}$Ca. It is instructive to note that the original calculation of Bahcall (1964) gave $\sigma(^8\text{B}) = 1.27(1.0 \pm 0.25) \times 10^{-42}$ cm$^2$ before any data on the decay of $^{37}$Ca was available and at a time when very little was known about the characteristics of the nuclear levels of $^{37}$Ar and $^{37}$K. The fact that the crudest original calculation agrees with the most refined existing calculation to within 13% gives us confidence in the validity of this important cross section.

An improved calculation of the neutrino scattering cross sections by Bahcall, Kamionkowski, and Sirlin (1995), which includes radiative corrections, has led to a net decrease in the total $\nu_e - e^-$ scattering cross sections by about 2%, essentially independent of energy. For $\nu_\mu - e^-$ scattering cross sections, the total cross sections are increased by about 1.3% relative to the values given in Bahcall (1989).

III. ELEMENT DIFFUSION

All of the detailed solar models published by different stellar evolution groups include modern input data for element abundances, nuclear cross sections, radiative opacities, equation of state, as well as the solar age, mass, and luminosity. However, as of this writing, very few stellar evolution include element diffusion. The reasons for this omission are, we believe, twofold: (1) element diffusion occurs on a long time scale (typically more than $10^{13}$ yr to diffuse a solar radius under solar conditions), which shows that the effects on stellar structure are small; and (2) diffusion mixes spatial and temporal derivatives, which causes complications in many stellar evolution codes. Nevertheless, the precision required for solar neutrino calculations and for evaluating helioseismological frequencies is so great that element diffusion influences significantly both the computed neutrino fluxes and the computed helioseismological frequencies (see Cox, Guzik, and Kidman, 1988; Bahcall and Loeb, 1990; Bahcall and Pinsonneault, 1992; Proffitt, 1994).

We have used in the present paper three different prescriptions to describe the effects of element diffusion: (1) an analytic treatment of helium diffusion (Bahcall and Loeb, 1990); (2) an accurate numerical treatment of helium diffusion (Thoul, Bahcall, and Loeb, 1994, TBL); and (3) an accurate numerical treatment of heavy element and helium diffusion (TBL).

In Bahcall and Pinsonneault (1992), we made use of the approximate analytic description of helium diffusion obtained by Bahcall and Loeb (1990). This prescription is used in the first five solar models we discuss in § IV (and Table III).

The main improvement in diffusion calculations that we introduce in this paper is to implement (see the last three models in Table III) the calculations of TBL, which are based upon exact numerical solutions by Thoul of the fundamental equations for element diffusion and heat transfer (the so-called Burgers' equations, Burgers, 1969). These solutions satisfy exactly the flow equations for a multicomponent fluid, including residual heat-flow terms. Unlike most previous discussions, no restriction is placed in Thoul's analysis upon the number of elements that can be considered nor upon their relative concentrations. The diffusion velocities are expected to be accurate to about 15% for solar conditions (TBL). Good agreement exists between the diffusion rates computed by Thoul *et al.* and the very different numerical treatment of Michaud and Proffitt (1993) (see also Proffitt, 1994).

An export subroutine that calculates the diffusion rates and which can be incorporated in stellar evolution codes has been developed by Thoul and is obtainable upon request (to JNB or A. Thoul). Moreover, Thoul *et al.* (1994) also provides simple analytic fits to the numerical results, which are convenient to implement within existing stellar evolution codes. In this paper, we use both the analytic fits and the exact numerical results.

In § IV, we first implement the Thoul *et al.* results for pure helium diffusion and show that the effects on the calculated neutrino fluxes are essentially identical to those obtained using the analytic approximations derived by Bahcall and Loeb (1990). We then include the effects of heavy element diffusion together with helium diffusion using both the analytic approximations and the export subroutine of TBL. The model containing heavy element diffusion according to the numerical calculations of TBL represents our best standard solar model.

All heavy elements were assumed to diffuse at the same rate as fully-ionized iron, which is a good approximation (see TBL) since the total effect of heavy element diffusion only causes a relatively small change in the stellar properties. For purposes of calculating the radiative opacity, the total $Z$ was adjusted only for diffusion; conversion in the CNO cycle was neglected. The change in the local CNO abundances was assumed proportional to the change in overall heavy element abundance. The conversion of CNO isotopes has only a small effect on the overall opacity (see, e. g., Paper I).

In implementing the diffusion of heavy elements, we have to change the radiative opacity at each spherical shell in the solar model after each time step. We calculate the effect of heavy element diffusion on the opacity by computing a total heavy element abundance at each solar radius, $Z(r)$, and then interpolating for the opacity between opacity tables with different total heavy element abundances. We performed a cubic spline interpolation in opacity as a function of $\rho/T_6^3$, $T$, and $Y$ for two different values of $Z$. We used linear interpolation in $Z$ between these two values to account for metal diffusion. Our principal approximation is to use the same Grevesse and Noels (1993a) relative mixture of heavy elements (see column two of Table II) at each point, although in fact different ions sink with slightly different velocities (see Figures 7 and 9 of TBL). We intend to improve this approximation in a future paper (Pinsonneault *et al.*, 1995, in preparation) in which we make use of tables to be calculated by Iglesias and Rogers which will describe the effect of each element on the total radiative opacity.

We make a small correction, $\lesssim 1\%$, to each neutrino flux to take account of the errors introduced by interpolation in $Z$ in the opacity table; this interpolation is required when metal diffusion is included. The correction was computed empirically by comparing two models run for the same conditions, both without metal diffusion. For one of the models, an opacity interpolation was artificially introduced, while the other model was computed using the exact values in the opacity table.

We estimate conservative uncertainties in the neutrino fluxes due to element diffusion in the following manner. We calculate each of the neutrino fluxes assuming that no diffusion (helium or metal) occurs and separately assuming both metal and helium diffusion occur at the rates given by the formalism of TBL. These calculations define a set of maximum fractional changes, $\alpha_i$, in the calculated values of each neutrino flux, $\phi_i$, according to the relation

$$\phi(\text{without diffusion})_i \equiv \phi(\text{with diffusion})_i(1 + \alpha_i). \tag{8}$$

The coefficients $\alpha_i$ are determined in § IV by comparing the neutrino fluxes for Model 9 and Model 10 of Table III. We follow TBL in estimating the $1\sigma$ intrinsic uncertainty in element diffusion to be

$$\Delta \alpha_i / \alpha_i = 0.15. \tag{9}$$

How far wrong could we possibly be in the estimated diffusion coefficients?

The separation of elements by diffusion can be inhibited by the presence of mixing in the solar interior. Although such mixing is not present in standard models, it is expected from some physical processes not usually included in standard stellar models, including turbulent diffusion (Schatzman, 1969; Lebreton and Maeder, 1986) and rotationally-induced mixing (Pinsonneault *et al.*, 1989; Pinsonneault, Kawaler, and Demarque, 1990; Chaboyer, Demarque, and Pinsonneault, 1995; see also Charbonnel, Vauclair, and Zahn, 1992). The strongest evidence for such mixing is the anomalous lithium depletion pattern seen in solar-type main sequence stars. In the sun, there is a factor of 160 difference in the meteoritic and photospheric lithium abundances.

Lithium is fragile, burning at a temperature of order $2.6 \times 10^6$ K. There is therefore a well-defined layer in stellar interiors below which lithium is destroyed. If the surface convection zone is deeper than this layer, surface lithium depletion occurs. On the main sequence the convection zone depth changes slowly with time, and stars of order 0.9 solar masses and higher are not expected to generate surface lithium depletion. During the pre-main sequence, higher mass stars can experience lithium depletion, but observations in young clusters such as the Pleiades (Soderblom *et al.*, 1993) place tight constraints on the allowed degree of pre-main sequence depletion. The stellar data shows evidence for main sequence lithium depletion (Hobbs and Pilachowski, 1988; Christensen-Dalsgaard, Gough, and Thompson, 1992; Pinsonneault and Balachandran, 1994; Balachandran, 1995). The time dependence of the depletion is not simply resolved by changes in the standard model physics because the depth of the convection zone is a strong function of mass, and lithium depletion occurs even in stars which are predicted to have very thin surface convection zones.

So what is the biggest error we could make? In our uncertainty estimates, we want to take account of the fact that there may be systematic uncertainties in the diffusion rate that have been considered. The most drastic assumption we can make is that the diffusion rates for helium and the heavy elements are all identically zero; this defines in the sense of Chapter 7 of Bahcall (1989) a strong $3\sigma$ lower limit to the effect of diffusion. This extreme limit applies if some unknown cause inhibits diffusion so that it does not occur at all. The asymmetry between the drastic lower-limit, no diffusion, and the smaller upper-limit (intrinsic) uncertainty, Equation (9), causes our uncertainty estimates for the neutrino fluxes and the predicted event rates in solar neutrino experiments to be asymmetric.

However, the survival of some of the lithium also places constraints on how much mixing can occur. Proffitt (1994) estimated that at most one-third of the effect of helium diffusion could be removed since some lithium survives. Using an approximate simultaneous treatment of helium diffusion and rotation, Chaboyer, Demarque, and Pinsonneault (1995) estimated that at most one-half of the effect of diffusion could be removed while preserving the observed amount of solar lithium. The exact degree of inhibition depends upon both the unknown time dependence and amplitude of the mixing and upon the unknown depth-dependence of the diffusion coefficients.

## IV. SOLAR MODELS

In this section, we calculate solar models using the improved input data described in § II and, for the first time in this series of papers, include the effects of metal diffusion. We proceed by a step-wise process: we make one improvement in the input data, calculate a new standard solar model, then make a further improvement in the input data and calculate a new standard model. The final improvement we make is to include metal diffusion according to the relatively accurate prescription of Thoul, Bahcall, and Loeb (1994).

In § IV.A, we describe a series of solar models in which each successive model contains one new improvement. The neutrino fluxes and the event rates in solar neutrino experiments are evaluated for each of the models in presented in Table III. We summarize the principal physical characteristics of each of these models in § IV.B and Table IV. In § IV.C, we present extensive numerical tables that describe the run of the physical variables within the sun, including the neutrino fluxes produced at each radial shell.

### A. Neutrino Fluxes in a Series of Solar Models

Table III gives the neutrino fluxes computed for each of the solar models.

The first model listed in Table III is the best solar model presented by Bahcall and Pinsonneault (1992), which includes helium diffusion and the recommended best values for all of the input data as of that writing. The second model in Table III shows the small decrease ($\sim 2\%$) in the predicted $^8$B neutrino flux and the $^{37}$Cl capture rate, that results when the radiative opacity for iron is calculated using intermediate coupling rather than $LS$-coupling (Iglesias, Rogers, and Wilson, 1992). For the third model, the Debye-Huckel correction to the pressure was made using the expression given by Cox and Giuli (1968), rather than the expression given by Bahcall and Pinsonneault (1992). The two expressions for the Debye-Huckel correction lead, as expected, to answers for the calculated neutrino fluxes that agree to within about 1%.

The improved solar luminosity and the more precise solar age are used in the next two models. The decrease in the adopted solar luminosity to the value given in § II.B decreases the calculated $^8$B neutrino flux and the chlorine event rate by about 3%. The use of the solar age given in § II.C decreases the $^8$B flux and the $^{37}$Cl capture rate by less than 1%. All the other changes in neutrino fluxes are even smaller.

The improved nuclear cross section factors that are described in § II.A and Table I make a somewhat more substantial change in the event rates. The sixth model in Table III shows that the improved nuclear data increases the calculated $^8$B flux and the $^{37}$Cl event rate by 6%, with smaller changes in the other entries in Table I.

The numerical treatment of helium diffusion by Thoul *et al.* (1994) (§ III) was implemented in the model represented in row seven of Table III. The net effect on the calculated neutrino fluxes of the numerical improvements is a little more than 4% compared to the analytic treatment of Bahcall and Loeb (1990).

The improved OPAL opacities (§ II.E) and the most recent heavy element abundances determined by Grevesse and Noels (1993a) (§ II.D) are used in constructing the model in row eight of Table III. The combined effect of these two changes is to decrease the $^7$Be flux by 2%, the $^8$B flux by about 4%, and the $^{37}$Cl and $^{71}$Ga event rates by about 4% and 1.5%, respectively.

The largest single change in the solar models is due to the inclusion of metal diffusion (cf. § III). The model that includes metal and helium diffusion as well as all of the other improvements mentioned above is shown in row nine of Table III. The inclusion of metal diffusion increases the $^7$Be neutrino flux by 6%, the $^8$B flux by 17%, the $^{37}$Cl event rate by 15%, and the $^{71}$Ga rate by 5%.

We determine the overall effect of hydrogen, helium, and heavy element diffusion by calculating a model with the same input parameters as used in our best model, row nine, but with no element diffusion (hydrogen, helium, or heavy elements). This No Diffusion model is presented in row ten. By comparing the results given in rows nine and ten we see that the overall effect of element diffusion is to increase the calculated $^7$Be flux by 13%, the $^8$B flux by 34%, the $^{37}$Cl event rate by 30%, and the $^{71}$Ga event rate by 8%.

B. Principal Physical Characteristics

Table IV presents some of the distinguishing physical characteristics of the computed solar models. The successive columns give the Model Name (same as in Table III), the central temperature and density, the heavy element to hydrogen ratio on the surface of the present sun, the initial mass fractions, the present central mass fractions, the ratio of the mixing-length to scale height, and the overall characteristics of the solar convective zone (the depth, the total mass, and the temperature at the base of the convective zone).

The series of physical characteristics given in Table IV can be used to evaluate the sensitivity of each of these characteristics to the set of input parameters that are varied in Table III and Table III. For example, the computed depth of the convective zone is different if one includes diffusion or neglects diffusion entirely. As can be seen from the third from last column of Table IV, the depth of the computed convective zone for our best model, which includes diffusion, is $R = 0.712\ R_\odot$, in agreement with the observed value determined from $p$-mode oscillation data of $R = 0.713 \pm 0.003\ R_\odot$ (Christensen-Dalsgaard *et al.*, 1991). On the other hand, if one neglects diffusion the depth of the convective zone is $R = 0.726$ $R_\odot$, in disagreement with the $p$-mode oscillation data.

Table V gives the computed surface properties of our best solar models with and without diffusion, Model 9 and Model 10 of Table III. The luminosity of the solar model increases by about 31% (or 32%) over the course of its main-sequence lifetime. Note that the depth of the convective zone is significantly deeper in the early solar lifetime, particularly in the model that includes diffusion.

C. Numerical Solar Models

Table VI and Table VII present, respectively, a detailed numerical description of the solar interior both with and without element diffusion. These details should be sufficient to permit accurate calculations of the effects of various proposed modifications of the weak interactions on the predicted neutrino fluxes. The first six columns of Table VI and Table VII give the physical variables that together help to define the model: the mass included in the current and all inner zones, the radius, the temperature (in degrees K), the density, the pressure, and the luminosity integrated up to and including the current zone. We use cgs units for the density and pressure. The last five columns of Table VI and Table VII give the principal isotopic abundances by mass, except for helium. The helium abundance is determined by the relation $Y = 1.0 - X - Z$, where the heavy element abundance $Z$ is given in Table IV.

Table VIII and Table IX present, for models with and without diffusion, the neutrino fluxes produced in a given spherical shell, as well as the temperature, electron number density, fraction of the solar mass, and the $^7$Be abundance by mass in the shell.

How do the gross surface properties of the model sun change with time? Table X gives the time-dependence calculated in models with and without element diffusion. The table shows how the calculated radius, luminosity, and depth of the convective zone evolve with time.

## V. UNCERTAINTIES

The interpretation of solar neutrino experiments in terms of either limits on the allowable range of solar interior characteristics or in terms of particle physics parameters requires quantitative estimates of the uncertainties in the predicted neutrino fluxes and neutrino interaction cross sections. Previous papers in this series (e.g., Bahcall *et al.*, 1982; Bahcall and Ulrich, 1988; and Bahcall and Pinsonneault, 1992) have described how the error estimates are derived. Historically, the theoretical uncertainty estimates played a major role in defining the solar neutrino problem since there was no other way to judge the seriousness of the discrepancy between theoretical solar calculations and experimental neutrino measurements (see, e.g., Bahcall and Davis, 1982).

Chapter 7 of Bahcall (1989) contains a systematic discussion of the procedures used to establish the uncertainties in the calculated neutrino fluxes and gives a large number of specific examples. We present here only a summary of the main ideas and present the specific results that follow from the uncertainties adopted in the present paper. Chapter 8 of Bahcall (1989) presents the calculated neutrino cross sections and their estimated uncertainties. We use these values except as explicitly noted in § II.F.

In this paper, we have finally succumbed to pressure from many of our experimental colleagues and are now quoting effective $1\sigma$ errors on all of the calculated quantities. We are not claiming that each of the error distributions is normally distributed with the errors given here, but rather that the $1\sigma$ values we quote are, for theoretical quantities, a reasonable approximation to what an experimentalist usually means by a $1\sigma$ error.

Previously, in this series of papers, we gave what we called "total theoretical errors", which translated into simple English meant we used $3\sigma$ input uncertainties for all measured quantities and, for theoretically-calculated quantities, the total range of published values to determine the (generally smaller) theoretical uncertainties. The $1\sigma$ theoretical uncertainties given here are equivalent to one-third the previously-quoted total theoretical errors. In some sense, we had no choice in this matter; our total theoretical uncertainties were already being divided by three and quoted as effective $1\sigma$ errors by experimentalists.

Table XI contains the calculated effective $1\sigma$ uncertainties in each of the principal solar neutrino fluxes from the ten most important input parameters. Columns three through seven of Table XI give the uncertainties in neutrino fluxes caused by uncertainties in the low-energy nuclear cross section factors. The last five columns give the uncertainties in the neutrino fluxes caused by lack of knowledge of the primordial heavy element to hydrogen ratio, the solar luminosity, the solar age, the radiative opacity, and the rate of element diffusion.

Nuclear-physics uncertainties are taken from the experimental papers and are summarized in Table I and in § II.A. Some authors have adopted larger errors than given in the published experimental nuclear physics papers, but the practice of using personal judgment to replace published errors removes much of the objective basis for the uncertainty estimates. The relatively small uncertainty, quoted here, in the cross section factor for the theoretically-calculated rate of the $p$-$p$ reaction is based upon the large amount of experimental data available for the nuclear two-body system and extensive numerical calculations of the allowed range of cross section factors that are consistent with this experimental data (see Kamionkowski and Bahcall, 1994a).

The uncertainty estimates for the heavy element to hydrogen ratio, the solar luminosity, and the solar age are determined in § II.B to § II.C of the present paper. The uncertainties in the solar radiative opacity were determined by comparing the results for accurate solar models computed with the older Los Alamos opacities and the much improved Livermore opacities. The effective $3\sigma$ uncertainty for the opacity is therefore,

$$\left[\frac{\Delta\phi}{\phi}\right]_{\text{opacity}} = 2\frac{[\phi(\text{Livermore}) - \phi(\text{Los Alamos})]}{[\phi(\text{Livermore}) + \phi(\text{Los Alamos})]} , \quad (10)$$

which spans the entire range between the old and the improved opacity calculations. Similarly, the $3\sigma$ lower-limit uncertainties in the neutrino fluxes caused by uncertainties in the diffusion rates are determined from the full differences in neutrino fluxes computed in two extreme models, the best No Diffusion Model 10 of Table III and the best model with diffu-sion (heavy element and helium diffusion), Model 9 of Table III. The specific prescriptions used to estimate the lower-limit and upper-limit uncertainties are given, respectively, in Equation 8 and Equation 9 of § III. The large uncertainty adopted here for the lower-limit to diffusion, requiring that the results overlap with the no-diffusion model at the $3\sigma$ lower-limit, causes the total theoretical uncertainties to be slightly larger in the present paper than in Bahcall and Pinsonneault (1992).

No single quantity dominates the uncertainties, given in Table III, in the individual neutrino fluxes. Therefore, it seems likely that—despite continuing efforts to improve the input parameters—the net uncertainties in the computed neutrino fluxes will not be greatly reduced in the foreseeable future.

## VI. SUMMARY AND DISCUSSION

We present and discuss in this section our principal conclusions.

(1) We have calculated improved values, or used recently-established improved values, for some of the most important input parameters for solar interior calculations and have determined best-estimates for their uncertainties. The parameters and their uncertainties are given in § II; the quantities considered include nuclear reaction rates, the solar luminosity, the solar age (see also Appendix A), heavy element abundances, radiative opacities, and neutrino interaction cross sections.

(2) The effects of the various improvements in the input parameters are determined systematically in § IV.A by calculating a series of accurate solar models, each model with an additional improvement in the input physics. The results of these calculations are summarized in Table III, which gives the neutrino fluxes for each model in the series, and in Table IV, which describes the principal physical characteristics of each model.

(3) We have included, for the first time in this series of papers, both heavy element and helium diffusion. We make use of an improved calculation of the diffusion coefficients by Thoul, Bahcall, and Loeb (1994), which is described in § III. The results obtained here for the neutrino fluxes are in good agreement with previous calculations by Bahcall and Pinsonneault (1992), if we restrict ourselves to only helium diffusion, and in good agreement with the results of Proffitt (1994) for the case in which metal diffusion is also included.

The inclusion of metal diffusion increases the calculated event rate in the chlorine experiment by about 12% (1.0 SNU), by about 5% (6 SNU) in the gallium experiments, by about 14% for the $^8$B neutrino flux (measured in the Kamiokande experiment), and by about 5% for the $^7$Be neutrino flux (to be measured in the BOREXINO experiment, see e.g., Ranucci, 1993).

For our best solar model with helium and heavy element diffusion, the predicted event rate for the chlorine experiment is $9.3^{+1.2}_{-1.4}$ SNU. For the gallium experiments, the predicted event rate is $137^{+8}_{-7}$ SNU. The model has a calculated flux of $5.15(1.00^{+0.06}_{-0.07}) \times 10^9$ cm$^{-2}$s$^{-1}$ for the $^7$Be neutrinos, and a flux of $6.6(1.00^{+0.14}_{-0.17}) \times 10^6$ cm$^{-2}$s$^{-1}$ for the $^8$B neutrinos, The uncertainties quoted here and elsewhere in this paper are effective $1\sigma$ uncertainties, calculated as described in § V.

The slightly higher predicted event rates found here for models that include heavy element diffusion only slightly exacerbate the difficulties in accounting for, with conventional physics, the observed neutrino event rates in the four existing experiments. Almost independent of the detailed results of solar models, it is difficult to explain the relative neutrino event rates in different detectors (see, e. g., Bahcall, 1994).

(4) The results obtained here by including metal diffusion and by using improved input parameters, increase the predicted event rates by about $1\sigma$ (theoretical) for the chlorine and the Kamiokande solar neutrino experiments and by about $0.5\sigma$ (theoretical) for the gallium experiments. Since all four of the operating experiments give rates that are lower than the predicted rates, one might suppose that the results given here make it more difficult to explain with conventional physics the solar neutrino results. However, the most obdurate difficulties are essentially independent of the details of the solar model physics; they result from comparisons between the experiments themselves (see, for example, Bahcall, 1994). We conclude, therefore, that the results presented here only slightly exacerbate the solar neutrino problem.

(5) Many authors have not yet included diffusion in their stellar evolution codes. In order to facilitate comparisons with their results, we have calculated a detailed solar model that uses all of the same physics as is used in our best solar except that no diffusion (by helium or heavy elements) is included. For this No Diffusion solar model (Model 10 in Table III), the predicted event rate for the chlorine experiment is $7.0^{+0.9}_{-1.0}$ SNU. For the gallium experiments, the predicted event rate is $126^{+6}_{-6}$ SNU. The model has a calculated flux of $4.5(1.00^{+0.06}_{-0.07}) \times 10^9$ cm$^{-2}$s$^{-1}$ for the $^7$Be neutrinos, and a flux of $4.9(1.00^{+0.14}_{-0.17}) \times 10^6$ cm$^{-2}$s$^{-1}$ for the $^8$B neutrinos,

Comparing the results obtained by including both metal and helium diffusion with the results obtained neglecting all diffusion (Model 9 and Model 10 in Table III), we find that the $^7$Be and $^8$B neutrino fluxes computed with models not including diffusion can be rescaled to take account of diffusion by multiplying the fluxes by, respectively, 1.14 and 1.36. The calculated event rates for the chlorine and gallium experiments can be rescaled to take account of the effects of diffusion by multiplying by 1.32 and 1.09, respectively. These ratios are useful when comparing the results obtained using solar models that do not take account of diffusion with results obtained by taking account of diffusion.

Rotationally-induced mixing may inhibit to some degree element diffusion. Indeed, some authors have argued that a moderate amount of inhibition may be required to explain the observed depletion of lithium in the solar atmosphere (see Chaboyer *et al.*, 1994 for a recent summary of these arguments). There is not yet available a rigorous calculation of the required amount of mixing that is independent of other uncertain parameters, such as the radiative opacity at the base of the solar convective zone. We regard the No Mixing model as an extreme example and have used this case in § V to determine effective $3\sigma$ lower-limits due to uncertainties in the diffusion rate.

(6) Models that include at least helium diffusion agree with the helioseismological determinations of the depth of the convective zone, while neglecting diffusion entirely leads to disagreement with the helioseismological data (see Table IV for details). The depth of the convective zone that is found when both metal and helium diffusion are included is $R = 0.712\ R_\odot$, in agreement with the observed value determined from $p$-mode oscillation data of $R = 0.713 \pm 0.003\ R_\odot$ (Christensen-Dalsgaard *et al.*, 1991). If only helium diffusion is included (Model 8 of Table III), the computed depth of the convection zone is $R = 0.710\ R_\odot$, which is within the quoted uncertainties of the $p$-mode determination. If diffusion is omitted entirely, the computed depth of the convective zone is $R = 0.726\ R_\odot$. Therefore, the no diffusion model disagrees with the $p$-mode data.

(7) Similarly, solar models must include at least helium diffusion in order to obtain agreement with the recent $p-$ mode determination of the surface abundance of helium, which yields $Y_s = 0.242 \pm 0.003$ (Hernandez and Christensen-Dalsgaard, 1994). The surface helium abundance that is found if just hydrogen and helium diffusion are included is $Y_s = 0.239$. If heavy element diffusion is also included, then $Y_s = 0.247$. These two solar models, full diffusion or just helium and hydrogen diffusion, yield surface helium abundances that bracket the observed value and are both within the measurement uncertainty. If diffusion is neglected entirely, the calculated surface abundance of helium, $Y_s = 0.268$, which disagrees with the observed value.

(8) The primordial helium abundance is determined to an accuracy of $\pm 2\%$ in our set of 10 solar models (see results in Table IV). For the full diffusion model (helium plus metal diffusion), the initial helium abundance is $Y_{\text{init}} = 0.278$. In our best no diffusion model, $Y_{\text{init}} = 0.270$. All 10 solar models discussed in this paper yield primordial helium abundances in the range $0.270 \leq Y \leq 0.278$.

(9) Diffusion causes the surface abundance of hydrogen to increase with time and the surface abundance of helium and the metals to decrease with time. Therefore, the heavy element abundance in the sun in the models with helium and metal diffusion, $Z_{\text{diffusion}} = 0.0200$, is about 14% larger than for models with out diffusion, $Z_{\text{no diffusion}} = 0.01740$. If helium but not metal diffusion is included, then $Z_{\text{He diffusion}} = 0.0182$ (cf. Table IV).

(10) Table XII gives for both our Best Model (Model 8 of Table III) and for the No Diffusion Model (Model 10 of Table III) the individual neutrino contributions to the calculated event rates in the chlorine and the gallium experiments. Our previous best solar model (see Model 1 of Table III or Bahcall and Pinsonneault, 1992), which included helium diffusion but not heavy element diffusion, predicts intermediate event rates, namely, 8.1 SNU for the chlorine experiment and 132 SNU for the gallium experiments.

(11) We present detailed numerical models for the physical characteristics of the sun calculated for our best solar model, including full helium and metal diffusion, and for an extreme model that neglects all diffusion. These numerical solar models are given in Table VI and Table VII; the corresponding neutrino production rates at different positions in the sun are given in Table VIII and Table IX. These tabular results can be used to evaluate the effects on the neutrinos of the solar material given a particular particle physics theory, such as the MSW theory. Copies of these tables may be obtained from the authors in computer readable form.

(12) The uncertainties in the different neutrino fluxes and calculated experimental event rates are evaluated quantitatively using the prescriptions summarized in § V. The results are presented in Table XI. No single input parameter dominates the estimated uncertainties.

The theoretical uncertainties are slightly increased because we have required that the $3\sigma$ lower-limits on the effects of diffusion include the results of the no-diffusion models. This new theoretical error affects most the lower limits on the calculated $^8$B neutrino fluxes and the neutrino capture rate in chlorine.

(13) Some workers (e.g., Hata, 1994; Wolfenstein, 1994) have drawn attention to the approximately $2\sigma$ difference between the two measurements of the $^7\text{Be}(p,\gamma)^8\text{B}$ low-energy cross sections that have the smallest quoted uncertainties. The reanalysis by Johnson *et al.* (1992) gives $25.2 \pm 2.4$ eV barn for the Kavanaugh (1969) measurement and $20.2 \pm 2.3$ eV barn for the Filippone *et al.* (1983) measurement. In the work in this paper, we have used the weighted average of five experimental values, including those of Kavanaugh (1969) and of Filippone *et al.* (1983), determined by Johnson *et al.* (1992) to be $22.4 \pm 2.1$ eV barn.

If we had used the cross section factor determined by Kavanaugh (1969) (or the cross section factor determined by Filippone *et al.*, 1983), the calculated event rate in the chlorine experiment would have been increased by 0.9 SNU or 10% (or decreased by −0.7 SNU or 8%), the calculated event rate in the gallium experiment would have been increased by 2.0 SNU or 1.5% (or decreased by 1%), the $^8$B neutrino flux would have been increased by 12.5% (or decreased by 10%), and the $^7$Be neutrino flux would have been unaffected. All of these changes are less than the quoted effective $1\sigma$ uncertainties.

ACKNOWLEDGMENTS

This work was supported in part by NSF grant PHY92-45317 at the Institute for Advanced Study. This work has been in progress for the past three years and has been discussed with many different colleagues; we are grateful to all of these colleagues for valuable comments, suggestions, and advice. A preliminary version of this work was described orally by JNB at the Neutrino 94 conference in Eilat, Israel in June 1994.

APPENDIX: AGE OF THE SOLAR SYSTEM

A lower bound to the age of the solar system is given by the age of material that has been melted and crystallized within the solar system. An upper bound to the age of the solar system is a time of significant injection of freshly synthesized presolar nucleosynthetic material into the proto-solar nebula. The "age" of the sun refers to the time since the protosolar mass arrived at some reference state in stellar evolution. Some objects in the solar system were formed as the result of melting of small planetary bodies, and some other objects appear to be melted and crystallized refractory condensates (CAI) formed in regions of the solar nebula that were at elevated temperatures. The melted refractory materials are known to have contained $^{26}$Al ($\bar{\tau} = 1.06 \times 10^6$ yr) at their time of formation (see Wasserburg, 1985 and Podosek and Swindle, 1988 for a review of short-lived nuclei in the early solar system). These objects also show clear isotopic differences due to nucleosynthetic processes for several elements as compared to terrestrial material, thereby demonstrating that they were formed from incompletely mixed presolar materials. The time interval between a melting and associated crystallization event and the present is determined by measurement of the number of parent nuclei that are left today and the increase in the number of daughter nuclei produced by the net decay of the parent nucleus over this time interval. The accuracy of the age depends on the measured abundances of each of the nuclear species, their initial abundances, and the decay constants. It is necessary that the samples measured have remained as isolated undisturbed systems since the time of melting (with isotopic homogenization) and associated crystallization and the present (cf. Wasserburg, 1987). The relations between the "age of the sun" and that of the planetary and solar nebular materials which have been dated depend on the sequence of formation and evolution of the sun and that of the planetary objects. The $^{26}$Al originally present in the CAI is considered to have been produced in presolar stellar sources. It thus follows that the ages of CAI place the strongest limits on the age of the sun, since the ratio of $^{26}$Al/$^{27}$Al in the nucleosynthetic processes in possible stellar sources can be well-bounded and compared with the observed values in CAI. The planetary objects are most plausibly formed after the sun but the corresponding stellar stage is not evident. There is a third class of materials, the chondrites, which are almost predominantly aggregates of processed and reprocessed (e.g., melted) nebular debris and possibly planetary materials. The chondrites have all undergone various degrees of recrystallization and frequently some degree of open system behavior; however, they also contain preserved presolar interstellar dust grains and thus have not been completely chemically reprocessed under either nebular or planetary conditions (cf. Black, 1972; Anders and Zinner, 1993). It is possible that $^3$He/$^4$He and D/H in gas-rich meteorites and the major planets may be used to establish bounds on the time of D burning and the formation of some planetary bodies (cf. Geiss and Reeves, 1981).

Samples of ancient nebular condensates and planetary materials represented by some meteorites are of sufficient mass to permit the application of a variety of isotopic dating methods. There are a large number of long-lived radioactive parent-daughter systems that have been used in dating meteorite samples. These methods include $^{238}$U–$^{206}$Pb, $^{235}$U–$^{207}$Pb, $^{232}$Th–$^{208}$Pb, $^{40}$K–$^{40}$Ar, $^{87}$Rb–$^{87}$Sr, $^{147}$Sm–$^{143}$Nd and $^{187}$Re $-^{187}$Os. Each of these systems have different susceptibilities to element redistribution and their decay constants ($\lambda_i$) have varying degrees of reliability. The inter-relationship between these chronometers plays a key role in establishing the evolution of the early solar system. An extensive review of the age of the solar system has been presented by Tilton (1988). There is generally good agreement between the different dating methods for systems with rather simple histories. For the purpose of focussing on the age of the sun using a self-consistent time scale with the best precision, we will concentrate on the $^{207}$Pb-$^{206}$Pb method. This assumes closed system behavior and utilizes the fact that the isotopic ratio of $^{238}$U/$^{235}$U is known and constant in solar system materials so that the isotopic ratio of radiogenic $^{207}$Pb to $^{206}$Pb (which can in many instances be determined precisely) can be used as a chronometer. The pair of lead isotopes provide model ages that are the most precise dating method available. The $^{207}$Pb-$^{206}$Pb model age uses the fact that these two isotopes of the element lead are produced by $^{235}$U and $^{238}$U respectively. The isotope $^{204}$Pb is unchanged over solar system history. The parent-daughter relationships are

$$^{207}\text{Pb}^* = \left[\left(\frac{^{207}\text{Pb}}{^{204}\text{Pb}}\right) - \left(\frac{^{207}\text{Pb}}{^{204}\text{Pb}}\right)_{\text{PAT}}\right]\left(^{204}\text{Pb}\right)_{\text{PAT}} = \ ^{235}\text{U}\left[e^{\lambda_{235}T} - 1\right] \qquad (\text{A1a})$$

and

$$^{206}\text{Pb}^* = \left[\left(\frac{^{206}\text{Pb}}{^{204}\text{Pb}}\right) - \left(\frac{^{206}\text{Pb}}{^{204}\text{Pb}}\right)_{\text{PAT}}\right]\left(^{204}\text{Pb}\right)_{\text{PAT}} = \ ^{238}\text{U}\left[e^{\lambda_{238}T} - 1\right] . \qquad (\text{A1b})$$

Here the asterisk refers to the number of radiogenic nuclei, $^i$Pb and $^i$U are the total number of $i$ nuclei in the sample and PAT refers to the primordial lead (Patterson, Brown, Tilton and Inghram, 1953). The ratio of these two equations ($^{207}$Pb*/$^{206}$Pb*) defines the $^{207}$Pb-$^{206}$Pb age and depends only on the lead abundances, the initial solar system Pb, and the ratio of $^{235}$U/$^{238}$U today. This method is resilient to any losses of lead or loss/addition of U in modern times. While problems exist with establishing closed-system behavior, there is often evidence of results self-consistent with $^{238}$U-$^{206}$Pb and $^{235}$U-$^{207}$Pb methods. The decay constants for $^{238}$U and $^{235}$U are also well established (Jaffey *et al.*, 1971), and the model ages are resilient to some types of open system behavior. This Pb-Pb method is one of the oldest dating methods used for solar system chronology (Nier, 1939; Houtermans, 1947). A broad and historical overview may be found in Dalrymple (1991). The development of sophisticated analytical techniques and superior sampling has permitted analyses of diverse meteorites and allowed substantial improvement in the results (cf. Wasserburg, 1987). In the following we will consider highly radiogenic Pb so that corrections for initial lead may be made with reliability. The isotopic composition of the initial solar system lead is known with good precision (Tatsumoto, Knight and Allègre, 1973; Chen and Wasserburg, 1983; Göpel, Manhès and Allègre, 1985).

Measurements of coarse-grained CAI give $^{207}$Pb/$^{206}$Pb model ages that are in essential agreement and lie within the range of $4.544 \pm 8$ to $4.565 \pm 5 \times 10^9$ yr (see, for example, Chen and Wasserburg, 1981). These samples also show that $^{26}$Al was present in them at the time they crystallized with a ratio $^{26}$Al/$^{27}$Al $= 5 \times 10^{-5}$ (Lee, Papanastassiou and Wasserburg, 1977). Assuming that $^{26}$Al/$^{27}$Al $= 1$ in the stellar source producing $^{26}$Al, we obtain for the time interval, $\Delta\tau$, between $^{26}$Al production and CAI formation $10.7 \times 10^6$ yr. The production ratio assumes complete conversion of $^{25}$Mg to $^{26}$Al with no destruction and no dilution. As such this value gives a maximum $\Delta\hat{\tau} = 10.7 \times 10^6$ yr. Decreasing the production ratio to 0.2 decrease $\Delta\tau$ by only $2 \times 10^6 y$. It follows that the age of the sun must be $t_{\odot\ \text{age}} < 4.581 \times 10^9$ yr old.

We now consider results on a meteorite (Angra Dos Reis) formed by melting and crystallization of a small planetary body (Angra Dos Reis Consortium, 1977). The resulting $^{207}$Pb $-^{206}$Pb ages range from 4.544 to $4.553 \times 10^9$ yr (Tatsumoto, Knight and Allègre, 1973; Wasserburg et al., 1977; Chen and Wasserburg, 1981). Another planetary differentiate (Ibitira) has been given precise $^{207}$Pb $-^{206}$Pb ages of $4.556 \pm 0.006$ (Chen and Wasserburg, 1985). In addition, we note that for two samples, the $^{238}$U $-^{206}$Pb$^{235}$U $-^{207}$Pb and $^{232}$Th $-^{208}$Pb ages are concordant within analytical errors ($\pm 0.05 \times 10^9$ yr). The minimum age for the sun must thus be $t_{\odot\ \text{age}} > 4.553 \times 10^9$ yr.

If we consider chondrites which are partially recrystallized aggregates, and focus on the ages of U-rich, $^{204}$Pb-poor phosphates, then the range in $^{207}$Pb $-^{206}$Pb analyzed by Chen and Wasserburg (1981) is $4.551 \pm 3$ and $4.552 \pm 4 \times 10^9$ yr. A thorough and more extensive study by Göpel et al. (1994) at higher precision, gives a range of $4.5044 \pm 5$ to $4.5627 \pm 7 \times 10^9$ yr. These ages reflect the times of formation of the phosphate grains that are considered to have grown in the meteorite after its accretion as a result of metamorphism and recrystallization. Assuming this to have occurred after the sun formed implies $t_{\odot\text{ age}} > 4.563 \times 10^9$ yr. A study of bulk chondrites by Unruh (1982) with corrections for contamination by terrestrial lead gives model $^{207}$Pb $-^{206}$Pb ages of $4.550 \pm 0.005 \times 10^9$ yr. These data appear in good agreement.

In conclusion, we infer from all of these results that the "age of the sun" is bounded by $4.563 \times 10^9$ yr $< t_{\odot\text{ age}} < 4.576 \times 10^9$ yr. The question of what stage of solar evolution this narrow time band represents is not at present obvious. It might be the very earliest phase or the start of main sequence behavior. If the $^{26}$Al that was present were produced by T-Tauri behavior of the early sun after blowing off of residual nebular gases, then the best age estimate would be $4.567 \times 10^9$ yr. However, we consider this to be an unlikely explanation for $^{26}$Al as it would not explain the presence of $^{107}$Pd (Kelly and Wasserburg, 1978), and for $^{53}$Mn (Birck and Allègre, 1985, 1988), it would yield values of $^{53}$Mn/$^{54}$Mn that were too high compared to $^{26}$Al/$^{27}$Al (Wasserburg and Arnould, 1987). If one considers ejecta from an Asymptotic Giant Branch star as the source of the short-lived species, $^{16}$Al, $^{107}$Pd and the recently discovered $^{60}$Fe in planetary bodies (Shukolyukov and Lugmair, 1993), then a dilution factor of $M_{\text{Heshell}}^{\text{AGB}}/M_{\text{ISM}} = 1.5 \times 10^4$ is determined and gives $\Delta \tau = 1 \times 10^6$ yr (Wasserburg et al., 1994). If the very short-lived $^{41}$Ca($\bar{\tau} = 1.5 \times 10^5$ yr) reported by Srinivasan, Ulyanov, and Goswami (1994) is from the same source, then $\Delta \tau = 0.5$ to $0.7 \times 10^6$ (Wasserburg et al., 1995). This would fix the age of the sun to be $t_{\odot} = 4.566 \pm 0.005 \times 10^9$ yr.

We note that this discussion has been directed to a restricted problem in order to obtain precise and self-consistent bounds to the "age of the sun."

# REFERENCES

Abdurashitov, J. N., *et al.*, 1994, Phys. Lett. B **328**, 234.

Akhmedov, E. Kh., 1988, Phys. Lett. B **213**, 64.

Anders, E., and N. Grevesse, 1989, Geochim. Cosmochim. Acta **53**, 197.

Anders, E., and E. Zinner, 1993, Meteoritics **28**, 490.

Angra Dos Reis Consortium, 1977, Earth Planet. Sci. Lett. **35**, 271.

Anselmann, P., *et al.*, 1994, Phys. Lett. B **327**, 377.

Aufderheide, M. B., S. B. Bloom, D. A. Resler, and C. D. Goodman, 1994, Phys. Rev. C **49**, 678.

Bahcall, J. N., 1964, Phys. Rev. Lett. **12**, 300.

Bahcall, J. N., 1989, *Neutrino Astrophysics* (Cambridge University Press, Cambridge, England).

Bahcall, J. N., 1994, Phys. Lett. B **338**, 276.

Bahcall, J. N., and R. Davis, Jr., 1982, in *Essays in Nuclear Astrophysics*, edited by C. A. Barnes, D. D. Clayton, and D. N. Schramm (Cambridge University Press, Cambridge, England), p. 243.

Bahcall, J. N., W. A. Fowler, I. Iben, and R. L. Sears, 1963, Astrophys. J. **137**, 344.

Bahcall, J. N., and A. Glasner, 1994, Astrophys. J. **437**, 485.

Bahcall, J. N., and B. Holstein, 1986, Phys. Rev. C **33**, 2121.

Bahcall, J. N., W. F. Huebner, S. H. Lubow, P. D. Parker, and R. K. Ulrich, 1982, Rev. Mod. Phys. **54**, 767.

Bahcall, J. N., M. Kamionkowski, and A. Sirlin, 1995, Phys. Rev. D (submitted).

Bahcall, J. N., and A. Loeb, 1990, Astrophys. J. **360**, 267.

Bahcall, J. N., and M. H. Pinsonneault, 1992, Rev. Mod. Phys. **64**, 885 (BP1992).

Bahcall, J. N., and R. K. Ulrich, 1971, Astrophys. J. **170**, 593.

Bahcall, J. N., and R. K. Ulrich, 1988, Rev. Mod. Phys. **60**, 297.

Balachandran, S., 1995, Astrophys. J., in press.

Birck, J.-L., and C. J. Allègre, 1985, Geophys. Res. Lett. **12**, 745.

Birck, J.-L., and C. J. Allègre, 1988, Nature **331**, 579.

Black, D. C., 1972, Geochim. Cosmochim. Acta **36**, 377.

Burgers, J. M., 1969, *Flow Equations for Composite Gases* (Academic Press, New York).

Castellani, V., S. Degl'Innocenti, G. Fiorentini, L. M. Lissia, and B. Ricci, 1994, Phys. Lett. B **324**, 425.

Chaboyer, B., P. Demarque, and M. H. Pinsonneault, 1994, Yale University preprint.

Chaboyer, B., P. Demarque, and M. H. Pinsonneault, 1995, Astrophys. J., in press.

Chapman, G. A., A. D. Herzog, J. K. Lawrence, S. R. Walton, H. S. Hudson, and B. M. Fisher, 1992, J. Geophys. Res. **97**, No. A6, 8211.

Charbonnel, C., S. Vauclair, and J.-P. Zahn, 1992, Astron. Astrophys. **255**, 191.

Chen, J. H., and G. J. Wasserburg, 1981, Earth Planet. Sci. Lett. **52**, 1.

Chen, J. H., and G. J. Wasserburg, 1983, Meteoritics **18**, 279.

Chen, J. H., and G. J. Wasserburg, 1985, Lunar Planet. Sci. Conf. XVI, p. 119.

Christensen-Dalsgaard, J., 1994, Europhysics News **25**, 71.

Christensen-Dalsgaard, J., D. O. Gough, and M. J. Thompson, 1991, Astrophys. J. **378**,

413.

Christensen-Dalsgaard, J., D. O. Gough, and M. J. Thompson, 1992, Astron. Astrophys. **264**, 518.

Christensen-Dalsgaard, J., C. R. Proffitt, and M. J. Thompson, 1993, Astrophys. J. Lett. **403**, 75.

Cox, J. P., and R. T. Giuli, 1968, in *Principles of Stellar Structure, Vol. 1, Physical Principles* (Gordon and Breach, New York), p. 345.

Cox, A., J. a. Guzik, and P. B. Kidman, 1989, Astrophys. J. **342**, 1187.

Dalrymple, G. Brent, 1991, *The Age of the Earth* (Stanford University Press, Stanford).

Davis, R., Jr., 1993, in *Frontiers of Neutrino Astrophysics*, edited by Y. Suzuki and K. Nakamura (Universal Academy Press, Inc., Tokyo), p. 47.

Dziembowski, W. A., P. R. Goode, A. A. Pamyatnykh, and R. Sienkiewicz, 1994, Astrophys. J. **432**, 417.

Elsworth, Y., R. Howe, G. R. Issak, C. P. McLeod, and R. New, 1990, Nature **347**, 536.

Filippone, B. W., A. J. Elwyn, C. N. Davids, and D. D. Koetke, 1983, Phys. Rev. C **28**, 2222.

Fröchlich, C., 1992, in *Proceedings of the Workshop of the Solar Electromagnetic Radiation Study for Solar Cycle 22*, edited by R. F. Donnelly (NOAA Space Environmental Laboratory, Boulder, CO),

Geiss, J., and H. Reeves, 1981, Astron. Astrophys. **93**, 189.

Göpel, C., M. Manhés, and C. J. Allègre, 1985, Geochim. Cosmochim. Acta **49**, 1681.

Göpel, C., M. Manhès, and C. J. Allègre, 1994, Earth Planet. Sci. Lett., in press.

Gough, D., and J. Toomre, 1991, Annu. Rev. Astron. Astrophys. **29**, 627.

Grevesse, N., 1984, Phys. Scripta **T8**, 49.

Grevesse, N., and A. Noels, 1993a, in *Origin and Evolution of the Elements*, ed. N. Prantzos, E. Vangioni-Flam, M. Cassé (Cambridge Univ. Press, Cambridge), p. 15.

Grevesse, N., and A. Noels, 1993b, Phys. Scripta **T47**, 133.

Gribov, V. N., and B. M. Pontecorvo, 1969, Phys. Lett. B **28**, 493.

Guenther, D. B., 1989, Astrophys. J. **339**, 1156.

Guenther, D. B., P. Demarque, Y.-C. Kim, and M. H. Pinsonneault, 1992, Astrophys. J. **387**, 372.

Guenther, D. B., M. H. Pinsonneault, and J. N. Bahcall, 1993, Astrophys. J. **418**, 469.

Guzzo, M. M., A. Masiero, and S. T. Petcov, 1991, Phys. Lett B **260**, 154.

Hata, N., 1994, private communication.

Hernandez, F. P., and J. Christensen-Dalsgaard, 1994, Mon. Not. R. Astron. Soc. **269**, 475.

Hickey, J. R., B. M. Alton, F. J. Griffin, H. Jacobowitz, P. Pellegrino, E. A. Smith, T. H. Vonder Haar, and R. H. Maschoff, 1982, J. Solar Energy **29**, 125.

Hickey, J. R., L. L. Stowe, H. Jacobowitz, P. Pellegrino, R. H. Maschoff, F. House, and T. H. Vonder Haar, 1980, Science **208**, 281.

Hirata, K. S., *et al.*, 1991, Phys. Rev. D **44**, 2241.

Hobbs, L. M., and C. A. Pilachowski, 1988, Astrophys. J. **334**, 734.

Houtermans, F. G., 1947, Das Alter des Urans. Z. Naturforschung **2A**, 322.

Hoyt, D. V., H. L. Kyle, J. R. Hickey, and R. H. Maschhoff, 1992, J. Geophys. Res. **97**, No. A1, 51.

Iben, I., and R. Talbot, 1966, Astrophys. J. **144**, 968.

Iglesias, C. A., and F. J. Rogers, 1991, Astrophys. J. **371**, 408.

Iglesias, C. A., F. J. Rogers, and B. G. Wilson, 1992, Astrophys. J. **397**, 717.

Jaffey, A. H., K. F. Flynn, L. E. Glendenin, W. C. Bentley, and A. M. Essling, 1971, Phys. Rev. C **4**, 1889.

Johnson, C. W., E. Kolbe, S. E. Koonin, and K. Langanke, 1992, Astrophys. J. **392**, 320.

Kamionkowski, M., and J. N. Bahcall, 1994a, Astrophys. J. **420**, 884.

Kamionkowski, M., and J. N. Bahcall, 1994b, Phys. Rev. C **49**, 545.

Kavanagh, R. W., T. A. Tombrello, J. M. Mosher, and D. R. Goosman, 1969, Bull. Am. Phys. Soc. **14**, 1209.

Kelly, W. R., and C. J. Wasserburg, 1978, Geophys. Res. Lett. **5** (12), 1079.

Kovetz, A., and G. Shaviv, 1994, Astrophys. J. **426**, 787.

Krishna Swamy, K. S., 1966, Astrophys. J. **145**, 174.

Kurucz, R. L., 1991, private communication.

Kurucz, R. L., 1991, in *Stellar Atmospheres: Beyond Classical Models*, edited by L. Crivellari, I. Hubeny, and D. G. Hummer (Kluwer, Dordrecht), p. 440.

Lambert, D. L., and B. Warner, 1968, Mon. Not. R. Astron. Soc. **140**, 197.

Langanke, K., 1994, in *Beyond the Standard Model III*, Proceedings of the INT-94-1 program on Solar Neutrinos and Neutrino Astrophysics (World Scientific, Singapore).

Lebreton, Y., and A. Maeder, 1987, Astron. Astrophys. **175**, 99.

Lee, R. B., M. A. Gibson, N. Shivakumar, R. Wilson, H. L. Kyle, and A. T. Mecherickunnel, 1991, Metrologia **28**, 265.

Lee, T., D. A. Papanastassiou, and G. J. Wasserburg, 1977, Astrophys. J. Lett. **211**, 107.

Libbrecht, K. G., 1988, Space Sci. Rev. **47**, 275.

Lim, C. S., and W. J. Marciano, 1988, Phys. Rev. D **37**, 1368.

Michaud, G. and C. R. Proffitt, 1993 in *Inside the Stars*, IAU Col. 137, edited by A. Baglin and W. W. Weiss (PASP, San Francisco), p. 246.

Mikheyev, S. P., and A. Yu. Smirnov, 1986, Sov. J. Nucl. Phys. **42**, 913; Nuovo Cimento **C9**, 17.

Nico, G., et al., 1995, in Proceedings of the XXVII International Conference on High Energy Physics, July 1994, Glasgow, eds. P. J. Bussey and I. G. Knowles (Institute of Physics: Philadelphia), p. 965.

Nier, A. O., 1939, Phys. Rev. **55**, 153.

Parker, P. D., 1994, in *Beyond the Standard Model III*, Proceedings of the INT-94-1 program on Solar Neutrinos and Neutrino Astrophysics (World Scientific, Singapore).

Patterson, C., H. Brown, G. Tilton, and M. Inghram, 1953, Phys. Rev. **92**, 1234.

Pinsonneault, M. H., and S. Balachandran, 1994, in *Eighth Cambridge Workshop on Cool Stars, Stellar Systems, and the Sun*, edited by J.-P. Caillault (ASP, San Francisco), **234**, 254.

Pinsonneault, M. H., S. D. Kawaler, and P. Demarque, 1990, Astrophys. J. Suppl. **74**, 501.

Pinsonneault, M. H., S. D. Kawaler, S. Sofia, and P. Demarque, 1989, Astrophys. J. **338**, 424.

Pinsonneault, M. H., *et al.*, 1995, in preparation.

Podosek, F. A.. and T. D. Swindle, 1988, *Meteorites and the Early Solar System*, edited by J. F. Kerridge and M. S. Matthews (Univ. of Arizona Press, Tucson), p. 1093.

Proffitt, C. R., 1994, Astrophys. J. **425**, 849.

Ranucci, G., 1993, for the Borexino Collaboration, Nucl. Phys. B (Proc. Suppl.) **32**, 149.

Rogers, F. J., and C. A. Iglesias, 1992, Astrophys. J. Suppl. **79**, 507.

Rogers, F. J., and C. A. Iglesias, 1994, Science **263**, 50.

Ross, J. E., and L. H. Aller, 1976, Science **191**, 1223.

Roulet, E., 1991, Phys. Rev. D **44**, R935.

Sackman, I.-J., A. I. Boothroyd, and W. A. Fowler, 1990, Astrophys. J. **360**, 727.

Schatzman, E., 1969, Astron. Astrophys. **3**, 331.

Shi, X., D. N. Schramm, and D. S. P. Dearborn, 1994, Phys. Rev. D **50**, 2414.

Shukolyukov, A., and G. W. Lugmair, 1993, Science **259**, 1138.

Soderblom, D. R., B. F. Jones, S. Balachandran, J. R. Stauffer, D. K. Duncan, S. B. Fedele, and J. D. Hudon, 1993, Astron. J. **106**, 1059.

Srinivasan, G., A. A. Ulyanov, and J. N. Goswami, 1994, Astrophys. J. Lett. **431**, 67.

Tatsumoto, M., R. J. Knight, and C. J. Allègre, 1973, Science **180**, 1279.

Thoul, A. A., J. N. Bahcall, and A. Loeb, 1994, Astrophys. J. **421**, 828.

Tilton, G. R., 1988, in *Meteorites and the Early Solar System*, edited by J. F. Kerridge and M. S. Matthews (University of Arizona Press, Tucson), p. 259.

Turck-Chièze, S., Cahen, S., Cassé, M., and Doom, C. 1988, Astrophys. J., 335, 415

Turck-Chièze, S., and I. Lopes, 1993, Astrophys. J. **408**, 347.

Unruh, D. M., 1982, Earth Planet. Sci. Lett. **58**, 75.

Wasserburg, G. J., 1985, in *Protostars and Planets II*, edited by D. C. Black and M. S. Matthews (University of Arizona Press, Tucson), p. 703.

Wasserburg, G. J., 1987, Earth Planet. Sci. Lett. **86**, 129.

Wasserburg, G. J., and M. Arnould, 1987, in *Nuclear Astrophysics*, edited by W. Hillebrandt, R. Kuhfuss, E. M. Aller, and J. W. Truran (Springer Verlag), p. 252.

Wasserburg, G. J., M. Busso, R. Gallino, and C. M. Raiteri, 1994, Astrophys. J. **424**, 412.

Wasserburg, G. J., R. Gallino, M. Busso, J. N. Goswami, and C. M. Raiteri, 1995, Astrophys. J., in press.

Wasserburg, G. J., D. A. Papanastassiou, and T. Lee, 1980, in *Early Solar System Processes and the Present Solar System*, (Corso Soc. Italiana di Fisica, Bologna).

Wasserburg, G. J., F. Tera, D. A. Papanastassiou, and J. C. Huneke, 1977, Earth Planet. Sci. Lett. **35**, 294.

Willson, R. C., 1993a, in *Atlas of Satellite Observations Related to Global Change*, edited by R. J. Gurney, J. L. Foster, and C. L. Parkinson (Cambridge University Press, New York), p. 5.

Willson, R. C., 1993b, private communication.

Willson, R. C., S. Gulkis, M. Janssen, H. S. Hudson, and G. D. Chapman, 1981, Science **211**, 700.

Wolfenstein, L., 1978, Phys. Rev. D **17**, 2369.

Wolfenstein, L., 1994, private communication.

TABLES

TABLE I.

TABLE II.

TABLE III.

TABLE IV.

TABLE V.

TABLE VI.

TABLE VII.

TABLE VIII.

TABLE IX.

TABLE X.

TABLE XI.

TABLE XII.